\documentclass[amsmath,amssymb,preprint,showpacs]{revtex4}

\usepackage{graphicx}
\usepackage[usenames]{color}

\begin{document}

\title{Temperature dependence of the excitation spectrum in the charge-density-wave ErTe$_3$ and HoTe$_3$ systems}

\author{F. Pfuner$^{1}$, P. Lerch$^{2}$, J.-H. Chu$^{3}$, H.-H. Kuo$^{3}$, I.R. Fisher$^{3}$, and L. Degiorgi$^{1}$}

\affiliation{$^{1}$Laboratorium f\"ur Festk\"orperphysik,
ETH-Z\"urich, CH-8093 Z\"urich, Switzerland\\
$^{2}$Swiss Light Source, Paul Scherrer Institute, CH-5232 Villigen-PSI, Switzerland\\
$^{3}$Geballe Laboratory for Advanced Materials
and Department of Applied Physics, Stanford University, Stanford, California
94305-4045, USA and Stanford Institute for Materials and Energy Sciences, SLAC National Accelerator Laboratory, 2575 Sand Hill Road, Menlo Park, California 94025, USA}

\date{\today}

\begin{abstract}
We provide optical reflectivity data collected over a broad spectral range and as a function of temperature on the ErTe$_3$ and HoTe$_3$ materials, which undergo two consecutive charge-density-wave (CDW) phase transitions at $T_{CDW1}$= 265 and 288 K and at $T_{CDW2}$= 157 and 110 K, respectively. We observe the temperature dependence of both the Drude component, due to the itinerant charge carriers, and the single-particle peak, ascribed to the charge-density-wave gap excitation. The CDW gap progressively opens while the metallic component gets narrow with decreasing temperature. An important fraction of the whole Fermi surface seems to be affected by the CDW phase transitions. It turns out that the temperature and the previously investigated pressure dependence of the most relevant CDW parameters share several common features and behaviors. Particularly, the order parameter of the CDW state is in general agreement with the predictions of the BCS theory.

\end{abstract}


\maketitle

\section{Introduction}
The charge-density-wave (CDW), first predicted by Peierls for one-dimensional interacting electron gas systems \cite{peierls}, is a prototype example of collective state, which, similarly to superconductivity, belongs to the class of broken-symmetry ground state. Two prominent energy scales characterize their electrodynamic response: the collective phason mode and the single-particle gap (2$\Delta$) excitation \cite{grunerbook}. The former mode is centered at zero frequency with vanishing width in the absorption spectrum of a superconductor, while in the CDW state the interaction between the condensate and the lattice imperfections shifts the mode to finite frequency, with the center (pinning) frequency usually well below the gap energy \cite{grunerbook}. For a superconductor, both collective and single particle excitations are possible and the relative strength of each depends on the ratio $\xi_0/l$, where $\xi_0$ is the coherence length and $l$ is the mean free path. On the other hand, for CDW's the collective mode spectral weight depends on $m_b/m^*$, where $m^*$ is the dynamical mass associated with the CDW and $m_b$ the band mass. The remaining spectral weight, 1-$m_b/m^*$, is associated with single particle transitions across the gap and, due to the prominent role played by the phonons ($m_b/m^*\le1$), most of the total spectral weight resides with the single particle excitations. Generally the two modes in the CDW state are well separated and each can be observed by measuring the conductivity $\sigma_1(\omega)$ over a broad spectral range. With transition temperatures on the order of 100 K, mean-field theory gives $2\Delta/h$ of the order of 100-1000 cm$^{-1}$, in the infrared range of frequencies, while the pinned mode resonance usually occurs in the millimeter wave spectral range in nominally pure specimens.

Another issue of general relevance in materials with strongly anisotropic properties concerns the role played by fluctuation effects \cite{schwartz}. The mean-field solution of an ideal one-dimensional system leads to a finite transition temperature $T^{MF}$ where long-range order develops and the system undergoes the Peierls transition to the CDW ground state. This, however, is an artifact of the mean-field approximation, which neglects the role played by fluctuations of the order parameter (i.e., the CDW gap). Real quasi-one-dimensional materials are highly anisotropic three-dimensional systems with interchain electronic Coulomb interactions and tunneling, leading to coupling of the fluctuations that develop along each chain. This coupling results in a finite transition temperature $T_{CDW}^{3D}$ below which three-dimensional long-range order occurs. For weak interchain coupling, $T_{CDW}^{3D}$ is significantly smaller than $T^{MF}$. The region below $T^{MF}$ is characterized by one-dimensional fluctuations which, at some temperature $T^*>T_{CDW}^{3D}$, cross over to fluctuations with two- or three-dimensional character. $T^*$ is the temperature at which the transverse correlation length $\xi_{\bot}$ becomes indeed comparable to the interchain spacing. The higher is the dimensionality of the interacting electron gas the smaller is the temperature interval where fluctuations occur.

These features pertaining to the CDW phase transition have been by now thoroughly explored and well established in the frequency dependent optical conductivity $\sigma(\omega)=\sigma_1(\omega)+i\sigma_2(\omega)$ of quasi one-dimensional (1D) chain-like materials \cite{grunerbook}. In 1D systems, the instability induced by the Peierls transition \cite{peierls,grunerbook} reflects furthermore a strong enhancement of the static electronic susceptibility, which develops at selected wavevectors spanning the Fermi surface (i.e., the so-called Fermi surface (FS) nesting at $q=2k_F$, $k_F$ being the Fermi wavevector). Evidences for the CDW state were also found in higher dimensions, specifically in novel two-dimensional (2D) layered compounds \cite{wilson,rouxel}. The 2D nature of the crystallographic structure leads to anisotropic physical properties and approximately cylindrical FSs. Obviously, the FS-driven CDW instabilities are generally weaker in 2D than in 1D systems. Nevertheless, under particular nesting conditions the electronic susceptibility can be sufficiently enhanced even in 2D for a CDW to develop. At variance with the 1D case however, 2D materials remain metallic in the presence of the CDW, since energy gaps can only open at discrete points of the FS. The persistence of such a metallic character in the broken-symmetry ground state also screens any signature of the collective mode in the electrodynamic response.

The understanding of layered correlated systems recently regained a lot of importance, because of the variety of correlations acting and of the instabilities occurring in 2D systems, the primary example of which is the high-temperature superconductivity in the cuprates. In this context, a family of layered compounds which have attracted a lot of attention recently are the rare-earth ($R$) tri-tellurides $R$Te$_3$, first studied by DiMasi et al.  \cite{dimasi}. Their crystallographic structure with space group $Cmcm$ \cite{norling} is made up of Te bilayers and insulating corrugated $R$Te slabs which act as charge reservoirs for the Te planes (inset Fig. 2a). $R$Te$_3$ host an $unidirectional$, incommensurate CDW already well above room temperature for all $R$ elements lighter than Dy \cite{ru1,ru2}, while in the heavy rare-earth tri-tellurides (i.e., $R$=Tm, Er, Ho, Dy) the corresponding transition temperature, $T_{CDW1}$, lies below $\sim$300 K and decreases with increasing $R$ mass. In the latter systems,  a further transition to a $bidirectional$ CDW state occurs at $T_{CDW2}$, ranging from 180 K for TmTe$_3$ to 50 K for DyTe$_3$ \cite{ru2}. Recent angle-resolved photoemission (ARPES) data \cite{moore} suggest that as the rare-earth ion is varied, the second CDW is formed only when the first CDW weakens with decreasing lattice parameters, making larger FS-segments available for the new nesting conditions to form \cite{ru2,Band2}. This second CDW modulation is along a direction perpendicular to the first one. The drastic change in transition temperatures $T_{CDW1}$ with the size of the $R$ ion or externally applied pressure on a given material \cite{sacchettiESRF} is accompanied by a similarly large change in the properties of the CDW itself. In particular, the CDW gap of the lighter $R$Te$_3$ was shown to progressively collapse when the lattice constant is equivalently reduced either by chemical or applied pressure, which, in turn, induces a transfer of spectral weight into the metallic component of the excitation spectrum  \cite{sacchetti1,sacchetti2,lavagniniprb}, the latter resulting from the fact that the FS in these 2D materials is indeed only partially gapped by the formation of the CDW. 

Here, we focus our attention on HoTe$_3$ and ErTe$_3$, as representative members of the heavy rare-earth tri-tellurides with transition temperatures into the uni- and bidirectional CDW states both below 300 K. This allows us to have access to the temperature dependence of the relevant energy scales (e.g., the CDW gaps), on which very little is known so far. From a broader perspective, we will compare the temperature dependence of the CDW phase transition with the previously studied \cite{sacchetti1,sacchetti2,lavagniniprb} impact of the lattice compression and of the pressure induced dimensionality crossover on the CDW formation. Furthermore, it is still widely debated to which extent fluctuations affect the CDW phase transition of the interacting electron gas in dimensions higher than one. Another goal of this work is thus to investigate the role played by the fluctuation effects with respect to the CDW phase transitions.

\begin{figure}[!tb]
\center
\includegraphics[width=9cm]{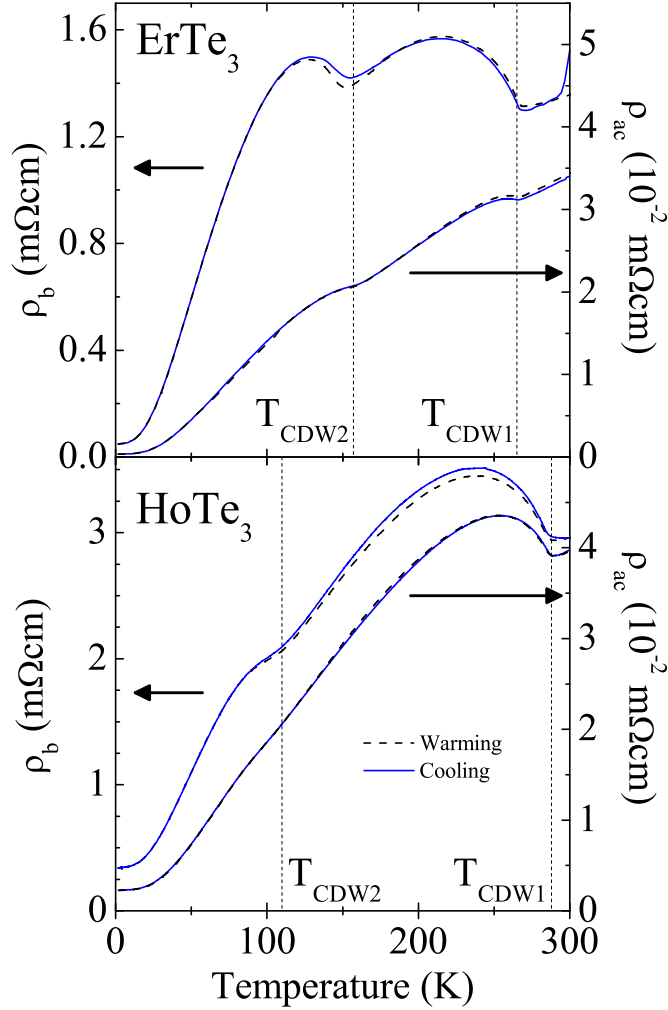}
\caption{(color online) Temperature dependence of the $dc$ resistivity for both title compounds within the $ac$ plane and along the orthogonal $b$ axis \cite{ru2}. Both warming (dashed lines) and cooling (solid lines) cycles are shown. The vertical dashed lines mark the critical temperatures at $T_{CDW1}$ and $T_{CDW2}$.}
\label{dc}
\end{figure}

\begin{figure}[!tb]
\center
\includegraphics[width=9cm]{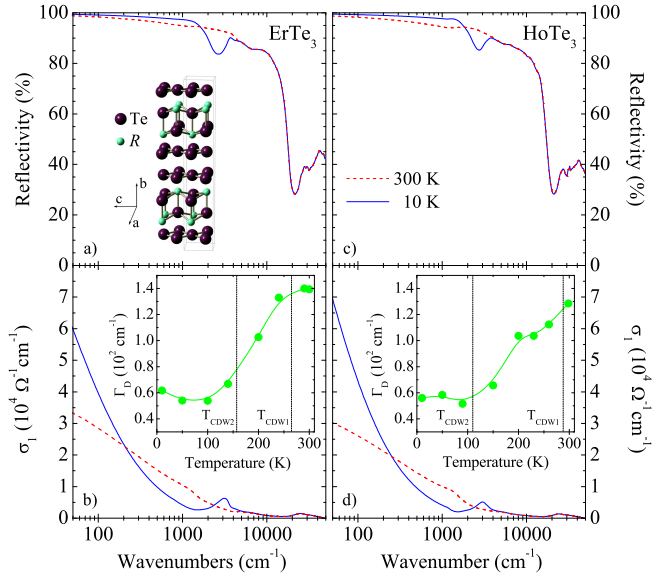}
\caption{(color online) (a-c) Optical reflectivity $R(\omega)$ of ErTe$_3$ and HoTe$_3$ at 10 and 300 K. The inset in panel (a) displays the crystal structure \cite{ru1}. (b-d) Real part $\sigma_1(\omega)$ of the optical conductivity of ErTe$_3$ and HoTe$_3$ at 10 and 300 K. The insets in panels (b-d) show the temperature dependence of the Drude scattering rate $\Gamma_D$ (see text). The vertical thin dotted lines mark the critical temperatures at $T_{CDW1}$ and $T_{CDW2}$.}
\label{Refl}
\end{figure}

\section{Experiment and Results}
The HoTe$_3$ and ErTe$_3$ single crystals were grown by slow cooling a binary melt, as described elsewhere \cite{ru1}. The plate-like crystals, removed from the melt by decanting in a centrifuge, could be readily cleaved between Te layers to reveal clean surfaces for the reflectivity measurements. All samples were thoroughly characterized with $dc$ transport method. Figure 1 displays the temperature dependence of the resistivity ($\rho_{dc}$) within the Te $ac$ plane and along the orthogonal $b$ axis \cite{ru2}. Distinct anomalies signal the CDW phase transitions at $T_{CDW1}$= 265 and 288 K and at $T_{CDW2}$= 157 and 110 K for ErTe$_3$ and HoTe$_3$, respectively. The overall trend of $\rho_{dc}(T)$ with decreasing temperature reinforces the notion that these materials remain metallic down to low temperatures and thus that only a fraction of the FS is affected by the CDW transitions.

We exploit several spectrometers and interferometers, in order to measure the optical reflectivity $R(\omega)$ as a function of temperature (4-300 K) for all samples from the far infrared (6 meV) up to the ultraviolet (6 eV) spectral range, with light polarized parallel to the Te planes. Dedicated optical cryostats with appropriate windows for the mid- and far-infrared range were employed. It is worth noting that data as a function of temperature in the mid-infrared range were collected with an infrared-microscope, which allowed us to investigate the optical response at several selected spots on the sample surface, therefore testing the degree of homogeneity of our samples. It turns out that our samples are very homogenous so that data collected on different spots are equivalent to each other. Details pertaining to the experimental methods can be found elsewhere \cite{Wooten,Dressel}. 

Figures 2a and 2c display $R(\omega)$ for both compounds at 300 K (i.e., in the so-called normal state) and at 10 K (i.e., within the CDW states with respect to both transitions at $T_{CDW1}$ and $T_{CDW2}$). The overall metallic character at any temperatures is well evident by the plasma edge feature at about 2x10$^4$ cm$^{-1}$ and by $R(\omega\rightarrow0)\rightarrow$ 100$\%$ (i.e., total reflection). At 10 K one can additionally recognize the depletion in $R(\omega)$ at about 3000 cm$^{-1}$. These features are very much reminiscent of what has been seen in our previous studies on $R$Te$_3$ as a function of chemical and applied pressure \cite{sacchetti1,sacchetti2,lavagniniprb}. In order to emphasize our findings on the temperature dependence of $R(\omega)$, we show in Fig. 3a and 4a the detailed view of $R(\omega,T)$ in the infrared range. For both compounds we observe the progressive development of the depletion at 3000 cm$^{-1}$ in $R(\omega)$ with decreasing temperature. As for the $R(\omega)$ data upon compressing the lattice, this feature will be later ascribed to the onset of the single particle excitation across the CDW gap. Its opening seems to continuously evolve with decreasing temperature, without any abrupt changes when crossing over from the undirectional CDW state below $T_{CDW1}$ to the bidirectional one at $T_{CDW2}$. Nevertheless, we do see a remarkable enhancement of $R(\omega)$ at low temperatures (i.e., $T\le$50 K), as consequence of the sharp onset and somehow steeper increase of $R(\omega)$ below the depletion at 3000 cm$^{-1}$ (Fig. 2a and 2c).

\begin{figure}[!tb]
\center
\includegraphics[width=8.2cm]{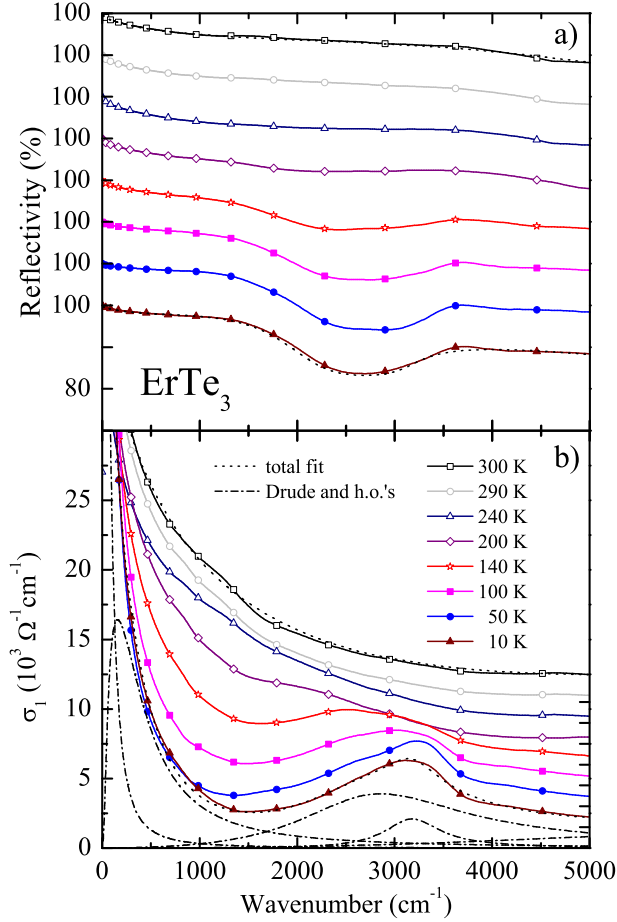}
\caption{(color online) Temperature dependence of $R(\omega)$ (a) and $\sigma_1(\omega)$ (b) in the infrared energy interval for ErTe$_3$. All spectra have been shifted for clarity by a constant value (in $R(\omega)$ by 10\% and in $\sigma_1(\omega)$ by 1.5x10$^{3}$ $(\Omega$cm)$^{-1}$). The thin dashed lines at 300 and 10 K in both panels correspond to the resulting total Lorentz-Drude fit (see text). In panel (b) the fit components are shown with dashed-dotted lines at 10 K.}
\label{Refl_sigmaEr}
\end{figure}

Exploring such a large spectral range allows us to perform reliable Kramers-Kronig (KK) transformations in order to achieve the optical functions. To this end, $R(\omega)$ was extended towards zero frequency (i.e., $\omega\rightarrow$0) with the Hagen-Rubens extrapolation ($R(\omega)=1-2\sqrt{\frac{\omega}{\sigma_{dc}}}$, $\sigma_{dc}$ being the $dc$ conductivity, Fig. 1 \cite{ru2}) and with standard power-laws (i.e., $R(\omega)\sim\omega^{-s}$, 2$\le s\le$4) at high frequencies \cite{Wooten,Dressel}. The overall view of the real part $\sigma_1(\omega)$ of the complex optical conductivity at two selected temperatures, above and well below $T_{CDW1}$ and $T_{CDW2}$, is shown for both compounds in Fig. 2b and 2d. Figures 3b and 4b then highlight $\sigma_1(\omega)$ in the infrared energy interval. Upon lowering the temperature we first observe a narrowing of the zero-energy  resonance, ascribed to the effective metallic (Drude) contribution to the absorption spectrum. Contrary to the 1D materials, where the CDW phase transition leads to an insulating state \cite{grunerbook}, the metallic part in $\sigma_1(\omega)$ of 2D materials overcasts the collective mode, making any attempts for its observation by infrared absorption methods impossible. Hand in hand with the narrowing of the metallic component there is the appearance of a pronounced mid-infrared feature, peaked at about 3000 cm$^{-1}$ and which obviously pairs with the depletion seen in $R(\omega)$ at about the same energy (Fig. 3a and 4a). The $\sigma_1(\omega)$ spectra as a function of temperature on HoTe$_3$ and ErTe$_3$ do share common features with our previous data upon lattice compression on the $R$Te$_3$ series \cite{sacchetti1, sacchetti2,lavagniniprb}. 

\begin{figure}[!tb]
\center
\includegraphics[width=8.2cm]{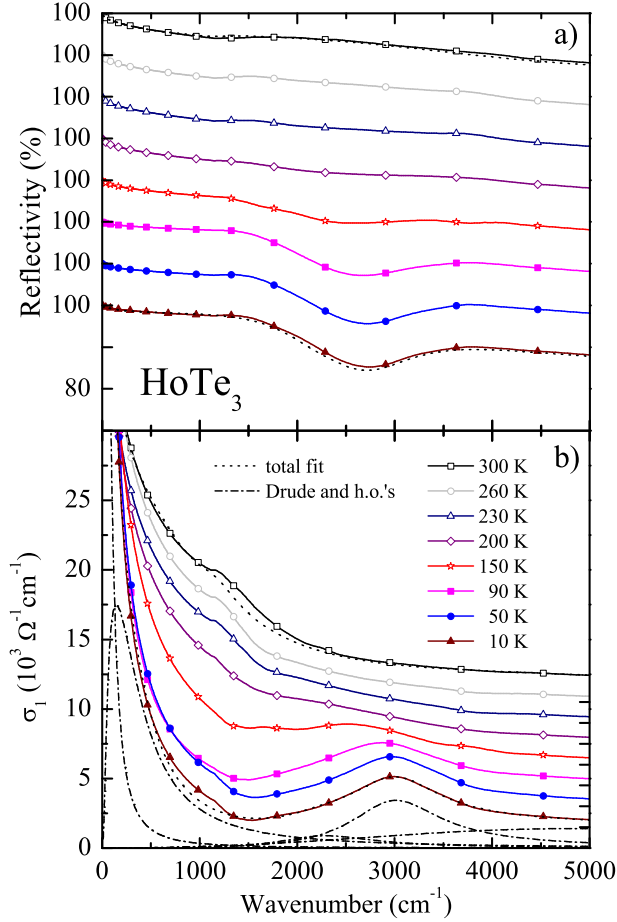}
\caption{(color online) Temperature dependence of $R(\omega)$ (a) and $\sigma_1(\omega)$ (b) in the infrared energy interval for HoTe$_3$. All spectra have been shifted for clarity by a constant value (in $R(\omega)$ by 10\% and in $\sigma_1(\omega)$ by 1.5x10$^{3}$ $(\Omega$cm)$^{-1}$). The thin dashed lines at 300 and 10 K in both panels correspond to the resulting total Lorentz-Drude fit (see text). In panel (b) the fit components are shown with dashed-dotted lines at 10 K.}
\label{Refl_sigmaHo}
\end{figure}

Higher energy excitations, corresponding to the
spectral features observed in $R(\omega)$ above the plasma edge,
are ascribed to electronic interband transitions. The frequencies
of these excitations are compatible with the predictions from
band-structure calculations \cite{Band2,Band}, taking into account
the band energies at the $\Gamma$-point of the Brillouin zone. On the other hand, those
band-structure calculations do not reveal any electronic
transition below 1 eV for the undistorted structure (i.e., in the
normal state). It is thus natural to identify the depletion in the
$\sigma_1(\omega)$ spectrum between the Drude and the mid-infrared
peak with the onset of the CDW gap. Therefore, the mid-infrared peak is
ascribed to the charge excitation across the CDW gap into a single
particle (SP) state. In the following we will refer to this feature
as the SP peak \cite{sacchetti1}. Upon increasing the temperature, the gap absorption progressively shifts into and is almost screened by the high-frequency tail of the metallic part in $\sigma_1(\omega)$ (Fig. 3b and 4b). Close to and above $T_{CDW1}$, the gap feature is indeed at best a broad shoulder overlapped to the Drude component.

\section{Discussion}
The temperature dependence of the two most prominent contributions to $\sigma_1(\omega)$, the effective metallic component and the CDW gap excitation, will be at the center of our discussion. To this end and in order to gain more insight into the absorption spectra we apply the standard Lorentz-Drude (LD) approach. It consists in reproducing
the dielectric function by the expression \cite{Wooten,Dressel}:
{\setlength\arraycolsep{2pt}
\begin{eqnarray}
\nonumber \tilde{\epsilon}(\omega) & = & \epsilon_1(\omega)
+i\epsilon_2(\omega) =
\\ & = & \epsilon_{\infty}-\frac{\omega_P^2}{\omega^2-i \omega
\Gamma_D}+\sum_j \frac{S_j^2}{\omega_j^2-\omega^2-i \omega
\gamma_j},
\end{eqnarray}}where $\epsilon_{\infty}$ is the optical dielectric constant,
$\omega_P$ and $\Gamma_D$ are the plasma frequency and the width
of the Drude peak, whereas $\omega_j$, $\gamma_j$, and $S^2_j$ are
the center-peak frequency, the width, and the mode strength for
the $j$-th Lorentz harmonic oscillator (h.o.), respectively.
$\sigma_1(\omega)$ is then obtained from $\sigma_1(\omega)=\omega
\epsilon_2(\omega)/4\pi$. The best fit to the data is achieved by a Drude term for the metallic contribution  and a selection of Lorentz harmonic oscillators (h.o.) for the finite frequency excitations. Figures 3b and 4b emphasize those fit components for the 10 K spectra. Applying such a fit procedure at all temperatures allows us to systematically reproduce the experimental data in great details, as illustrated for both $R(\omega)$ and $\sigma_1(\omega)$ at 300 and 10 K in Fig. 3 and 4. Similarly to our previous optical work as a function of pressure \cite{sacchetti1,sacchetti2,lavagniniprb}, we can extract from the fits several characteristic energy scales and optical parameters, which can be related to the CDW phase transitions. The emphasis will be here on the (Drude) scattering rate, the CDW gap and the fraction of the ungapped FS. 

As directly evinced from the data, the metallic component in $\sigma_1(\omega)$ narrows with decreasing temperature. This is well represented by the Drude scattering rate $\Gamma_D$, shown in the insets of Fig. 2b and 2d as a function of temperature for both compounds. $\Gamma_D$ rapidly decreases with decreasing temperature, indicating that deep into the CDW state less scattering channels seem to be available. This nicely correlates with the overall trend observed in the $dc$ transport properties (Fig. 1) \cite{ru2}. The metallic component in $\sigma_1(\omega)$ is nonetheless characterized by a pronounced high frequency tail which deviates from a purely Drude shape (Fig. 3b and 4b). This may suggest a scenario for the conduction band consisting of itinerant (Drude) charge carriers close to the Fermi level and more localized carriers below the mobility edge. Therefore, one can extend the description of the metallic part in $\sigma_1(\omega)$, based on the single Drude term, to an approach given by the combination of both Drude term and first Lorentz h.o. (Fig. 3b and 4b). Consequently, the alternative way to extract the scattering rate would be to invert the optical conductivity within the generalized Drude model, encountering the spectral weight in $\sigma_1(\omega)$ up to a cut-off frequency of $\sim$ 1500 cm$^{-1}$ \cite{Dressel}. One can first obtain the frequency dependence of the scattering rate $\Gamma(\omega)$ and consequently its static $(\omega\rightarrow 0)$ limit. It turns out that $\Gamma(\omega\rightarrow 0)$ as well as $\Gamma_D$ display the same relative temperature dependence, thus reinforcing our conclusions based on the trend in temperature of the transport properties.

\begin{figure}[!tb]
\center
\includegraphics[width=8.2cm]{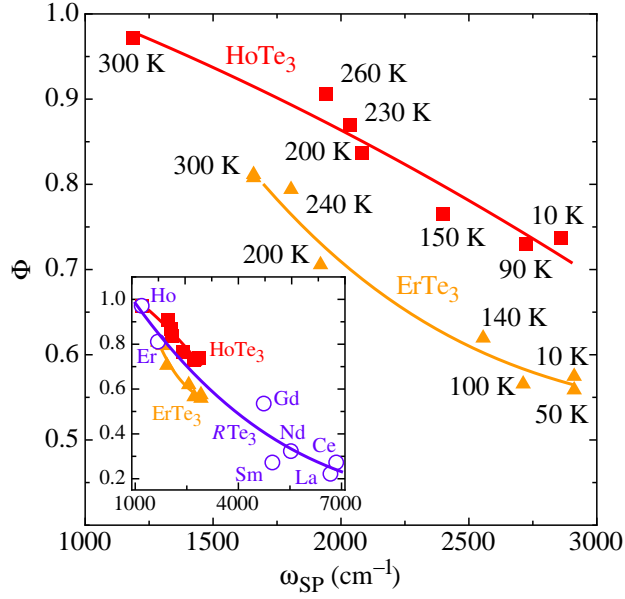}
\caption{(color online) The ratio $\Phi$ of the ungapped Fermi surface plotted versus the single particle excitation $\omega_{SP}$ (i.e., CDW gap) for both title compounds. Temperature is here an implicit variable. The inset compares the trend of $\Phi$ versus $\omega_{SP}$ in temperature (from the main panel) for ErTe$_3$ and HoTe$_3$ with the chemical pressure results for selected compounds of the $R$Te$_3$ series \cite{sacchetti1}. The polynomial lines through the data in main panel and inset are meant as guide to the eyes.}
\label{param}
\end{figure}

Since the first Lorentz h.o. merges into the Drude tail and should count as part of the effective metallic contribution to $\sigma_1(\omega)$, the two Lorentz h.o.'s, covering the range between 2000 and 5000 cm$^{-1}$ (Fig. 3b and 4b), then define the average weighted energy scale $\omega_{SP}$:
\begin{equation}
\omega_{SP}=\frac{\sum_{j=2}^3 \omega_j S_j^2}{\sum_{j=2}^3
S_j^2},
\end{equation}
which we associate with the CDW gap excitation (i.e., $\omega_{SP}=2\Delta$) \cite{comment}. Within the same phenomenological approach adopted here for the estimation of $\omega_{SP}$, we can also estimate the fraction of the FS, which then remains ungapped across the CDW phase transitions, as follows \cite{sacchetti1}:
\begin{equation}
\Phi=\frac{\omega_p^2+S_1^2}{(\omega_p^2+\sum_{j=1}^3 S_j^2)}.
\end{equation}
Figure 5 displays $\Phi$ versus $\omega_{SP}$, the temperature being here an implicit variable. 
Upon increasing the temperature, we observe that the smaller is the CDW gap the larger is the fraction of the ungapped FS. This is totally in accordance with our previous investigation upon compressing the lattice \cite{sacchetti1,sacchetti2,lavagniniprb}, which reveals the simultaneous closing of the CDW gap with the enhancement of $\Phi$. This is emphasized in the inset of Fig. 5, where the implicit temperature dependence of $\Phi$ versus $\omega_{SP}$ for the title compounds is in trend with the behavior given by the chemical pressure when going from the lighter to the heavier $R$Te$_3$ \cite{sacchetti1}. About 30-40$\%$ of FS in ErTe$_3$ and HoTe$_3$ turns out to be affected at low temperatures by the formation of the CDW condensates. This is consistent with the tight-binding (TB) estimation of the density of states at the Fermi level, which seems to be suppressed to $\sim$ 77 \% of the unmodulated value by the first CDW transition and further to $\sim$ 74 \% by the second one \cite{brouet}. $\Phi$ undergoes minor changes at temperatures below $T_{CDW2}$ (Fig. 5), supporting the notion evinced from TB that the gains due to the second CDW are modest.

We now turn our attention to the explicit temperature dependence of $\omega_{SP}$, which is shown in Fig. 6. $\omega_{SP}(T)$ for HoTe$_3$ and ErTe$_3$ is here normalized by its value deep into the CDW ground state (i.e., $\omega_{SP}$(10 K)), while the temperature axis is normalized by the respective $T_{CDW1}$. As expected, $\omega_{SP}$ monotonically increases with decreasing temperature below $T_{CDW1}$. It is worth noting that the optical estimation of the CDW gap for ErTe$_3$ at 10 K (i.e., $\omega_{SP}\sim$ 3000 cm$^{-1}$, Fig. 5) is nearly identical with the gap value extracted from ARPES experiment \cite{moore}. However, the temperature dependence of $\omega_{SP}$ does not display any clear-cut anomalies in coincidence with the second phase transition at $T_{CDW2}$. On the contrary, the ARPES investigation does give evidence for a smaller gap (i.e., of the order of 800 cm$^{-1}$) due to the second CDW \cite{moore}. Nevertheless, ARPES also establishes that the area of FS gapped by the first CDW transition is three times the area gapped by the second one \cite{moore}. Therefore, the excitation due to the smaller gap is most probably overcast by the effective metallic contribution in the absorption spectrum and could well merge into the high frequency tail of the Drude resonance, as emphasized above. Moreover, with our optical method we only measure the average energy-excitation over the whole Brillouin zone, missing the $k$-space resolution of ARPES. 

A signature of the gap feature is already present at 300 K, close to but yet above the first high temperature CDW phase transition at $T_{CDW1}$ (Fig. 6). We remark that this is a rather common situation in prototype CDW materials \cite{schwartz}. The persistence of the gap above the phase transition temperature can be considered as a fingerprint of precursor effects of the CDW formation and has been widely invoked as a manifestation of the fluctuation regime \cite{schwartz}. Therefore, CDW fluctuations seem to play an important role in $R$Te$_3$ as well, despite their two-dimensionality. The presence of substantial fluctuations is also confirmed by the observation of superlattice peaks with rapidly increasing width and decreasing correlation length well above $T_{CDW1}$ \cite{ru2}.

\begin{figure}[!tb]
\center
\includegraphics[width=8.2cm]{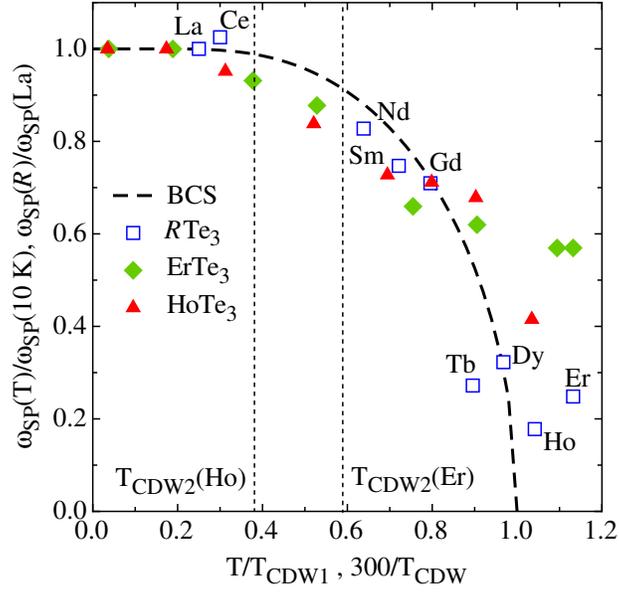}
\caption{(color online) Temperature and chemical pressure dependence of $\omega_{SP}$, normalized by the low temperature values of ErTe$_3$ and HoTe$_3$ or by the values of LaTe$_3$ in the $R$Te$_3$ series. The temperature axis is normalized by the respective critical temperatures ($T_{CDW1}$ for ErTe$_3$ and HoTe$_3$, or $T_{CDW}$ for the $R$Te$_3$ series, see text) \cite{ru1,ru2}. The vertical thin dotted lines mark the critical temperatures $T_{CDW2}$ for ErTe$_3$ and HoTe$_3$. The BCS predictions \cite{tinkham} for the order parameter is shown for comparison.}
\label{merit}
\end{figure}

It is now instructive to compare the relevant parameter $\omega_{SP}$ achieved with our optical experiments when varying the temperature (this work) and upon lattice compression \cite{sacchetti1}. For the purpose of clarity, we limit our comparison to the chemical series, having already showed that both chemical and applied pressure are equivalent \cite{sacchetti1,sacchetti2,lavagniniprb}. Figure 6 additionally displays the gap ratio for the chemical series (open squares) \cite{sacchetti1}. We took the $\omega_{SP}$ values measured at 300 K, normalized with the gap of LaTe$_3$, assumed to be the largest one for $R$Te$_3$. Instead of the normalized temperature, we consider here the ratio 300 K/T$_{CDW}$ as the effective temperature axis for the chemical series. In fact, it has been shown both with optics and ARPES \cite{sacchetti1,brouet} that upon chemically compressing the lattice, the shape of FS changes to such an extent that the ideal nesting conditions are progressively suppressed. This leads to a reduction of $T_{CDW}$, which are the measured CDW critical temperatures extracted from the $dc$ transport data of $R$Te$_3$ \cite{ru1,ru2}. Therefore, reducing the lattice constant along the chemical series at constant room temperature may be considered to be equivalent as changing the relative temperature difference between 300 K (where the measurement is performed) and $T_{CDW}$. Within the CDW state, we evince a general common trend in the development of the gap for the $R$Te$_3$ compounds, when changing $R$, as well as for HoTe$_3$ and ErTe$_3$, when varying the temperature. As comparison, we reproduce the BCS temperature dependence of the order parameter \cite{tinkham}.

The resulting overall decrease of the CDW gap equivalently with increasing temperature or lattice compression roughly agrees with the theoretical predictions, based on the mean-field like BCS theory. Deviations might be related to the uncertainties implicit in our fitting procedure towards the estimation of $\omega_{SP}$, which cannot be excluded a priori, although very unlikely. On the contrary at least as far as the temperature dependence is concerned, we tend to believe that our comparison with respect to a single BCS order parameter is too crude, since it neglects the cascade-like onset of the two CDW transitions in ErTe$_3$ and HoTe$_3$. Indeed in the temperature interval around $T_{CDW2}$ one encounters the largest deviation from the BCS calculations. Interestingly enough, the recent ARPES investigation also pointed out that, while the closing of the gaps is suggestive of a mean-field type behavior, the temperature dependence of the larger CDW gap is somewhat suppressed from the mean-field curve \cite{moore}. Our recent Raman scattering study \cite{lavagninicondmat} further emphasized the complex behavior of the collective CDW state within the unidirectional phase as well as at the crossover from the uni- to the bidirectional CDW state. Supported by first principles calculation \cite{lavagniniRaman} we reconcile the unconventional collective signatures of the CDW states with the bilayer structure (inset Fig. 2a) of the compounds under investigation. Nevertheless, a comprehensive theoretical approach, addressing the interplay of both CDW phase transitions, is still missing. 

\section{Conclusions}
We have studied the temperature dependence of the optical properties of the layered two-dimensional HoTe$_3$ and ErTe$_3$ compounds, which undergo two CDW phase transitions into a unidirectional and bidirectional state. From the absorption spectrum, we have extracted the temperature dependence of the order parameter, which gives clear-cut evidence for fluctuation effects. Moreover, we estimated the fraction of FS, which is affected by the formation of the CDW collective states. Our findings generally agree with the BCS predictions and reinforce the notion that increasing temperature is rather equivalent to lattice compression when destroying the CDW state. Similar to conclusions drawn from recent ARPES results on the same compounds, our optical data also foresee a dynamical interplay of the two CDWs, a theoretical description of which is highly desired.

\begin{acknowledgments}
The authors wish to thank A. Kuzmenko for technical support, and M. Lavagnini and R. Hackl for fruitful
discussions. This work has been
supported by the Swiss National Foundation for the Scientific
Research as well as by the NCCR MaNEP pool. Work at Stanford is supported by the Department of
Energy, Office of Basic Energy Sciences under contract
DE-AC02-76SF00515.
\end{acknowledgments}


\begin{thebibliography}{99}


\bibitem{peierls} R. Peierls, \emph{Quantum Theory of Solids}, Clarendon Press,
   Oxford (1955).

\bibitem{grunerbook} G. Gr\"uner, \emph{Density Waves in Solids}, Addison
   Wesley, Reading, MA (1994).

\bibitem{schwartz} A. Schwartz, M. Dressel, B. Alavi, A. Blank, S. Dubois, G. Gr\"uner, B. P. Gorshunov, A. A. Volkov, G. V. Kozlov, S. Thieme, L. Degiorgi 	and F. Levy, \emph{Phys. Rev. B} \textbf{52}, 5643 (1995).

\bibitem{wilson} J.A. Wilson, F.J. Di Salvo and S. Mahajan,
 \emph{Adv. Phys.} \textbf{24}, 117 (1975).

\bibitem{rouxel}
J. Rouxel, in \emph{Crystal Chemistry and Properties of Materials with quasi-one-dimensional Structures}, Eds. J. Rouxel and D. Riedel, Dordrecht (1986), pp. 1-26.

\bibitem{dimasi} E. DiMasi, M.C. Aronson, J.F. Mansfield, B. Foran and S. Lee,
 \emph{Phys. Rev. B} \textbf{52}, 14516 (1995).
 
 \bibitem{norling} B.K. Norling and H. Steinfink, \emph{Inorg. Chem.} \textbf{5}, 1488 (1966).
 
\bibitem{ru1} N. Ru and I.R. Fisher, \emph{Phys. Rev. B} \textbf{73},
   033101 (2006).

\bibitem{ru2} N. Ru, C.L. Condron, G.Y. Margulis, K.Y. Shin, J. Laverock, S.B. Dugdale, M.F. Toney and I.R. Fisher,
 \emph{Phys. Rev. B} \textbf{77}, 035114 (2008).

\bibitem{moore} R.G. Moore, V. Brouet, R. He, D.H. Lu, N. Ru, J.-H. Chu, I.R. Fisher and Z.-X. Shen,  \emph{Phys. Rev. B} \textbf{81}, 073102 (2010).

\bibitem{Band2} J. Laverock, S.B. Dugdale, Zs. Major, M.A. Alam, N. Ru, I.R. Fisher, G. Santi and E. Bruno, 
\emph{Phys. Rev. B} \textbf{71}, 085114 (2005).

\bibitem{sacchettiESRF} A. Sacchetti, C.L. Condron, S.N. Gvasaliya, F. Pfuner, M. Lavagnini, M. Baldini, M.F. Toney, M. Merlini, M. Hanfland, J. Mesot, J.-H. Chu, I.R. Fisher, P. Postorino and L. Degiorgi, \emph{Phys. Rev. B} \textbf{79}, 201101(R) (2009).

\bibitem{sacchetti1} A. Sacchetti, L. Degiorgi, T. Giamarchi, N. Ru and I.R. Fisher, 
\emph{Phys. Rev. B} \textbf{74}, 125115 (2006).

\bibitem{sacchetti2} A. Sacchetti, E. Arcangeletti, A. Perucchi, L. Baldassarre, P. Postorino, S. Lupi, N. Ru, I.R. Fisher and L. Degiorgi,
 \emph{Phys. Rev. Lett.} \textbf{98}, 026401 (2007).

\bibitem{lavagniniprb} M. Lavagnini, A. Sacchetti, C. Marini, M. Valentini, R. Sopracase, A. Perucchi, P. Postorino, S. Lupi, J.-H. Chu, I.R. Fisher and L. Degiorgi, \emph{Phys. Rev. B} \textbf{79}, 075117 (2009).

\bibitem{Wooten} F. Wooten, \emph{Optical Properties of Solids},
Academic Press, New York (1972).

\bibitem{Dressel} M. Dressel and G. Gr\"uner, {\itshape
Electrodynamics of Solids}, Cambridge University Press (2002).

\bibitem{Band} A. Kikuchi, \emph{J. Phys. Soc. Jpn.} \textbf{67}, 1308
(1998).

\bibitem{comment} This is at variance to our previous analysis \cite{sacchetti1}, where the first three Lorentz h.o.'s were associated with $\omega_{SP}$. Nevertheless, the overall trend of $\omega_{SP}$ for $R$Te$_3$ is also confirmed by reanalyzing the data on the chemical series with the present new fit procedure. 

\bibitem{brouet} V. Brouet, W.L. Yang, X.J. Zhou, Z. Hussain, R.G. Moore, R. He, D.H. Lu, Z.X. Shen, J. Laverock, S.B. Dugdale, N. Ru and I.R. Fisher, \emph{Phys. Rev. B}
\textbf{77}, 235104 (2008).

\bibitem{tinkham} M. Tinkham, {\em Introduction to superconductivity}, $2^{nd}$ Ed., McGraw-Hill, New York (1996).

\bibitem{lavagninicondmat} M. Lavagnini, H.-M. Eiter, L. Tassini, B. Muschler, R. Hackl, R. Monnier, J.-H. Chu, I.R. Fisher and L. Degiorgi, \emph{Phys. Rev. B} \textbf{81}, 081101(R) (2010).

\bibitem{lavagniniRaman} M. Lavagnini, M. Baldini, A. Sacchetti, D. Di Castro, B. Delley, R. Monnier, J.H. Chu, N. Ru, I.R. Fisher, P. Postorino and L. Degiorgi, \emph{Phys. Rev. B} \textbf{78}, 201101(R) (2008).

\end{thebibliography}
\end{document}